\documentstyle[pre, aps, twocolumn, epsfig]{revtex}

\begin{document}
\draft

 \twocolumn[\hsize\textwidth\columnwidth\hsize\csname@twocolumnfalse\endcsname

\title{Interaction Round a Face DMRG method \\
applied to rotational invariant quantum spin chains}
\author{Wada Tatsuaki \footnote{e-mail address: wada@ee.ibaraki.ac.jp} }
\address{Department of Electrical and Electronic Engineering, Ibaraki 
University, Hitachi, 316-8511, Japan}
\maketitle

\begin{abstract}
An interaction-round-a-face density-matrix renormalization-group 
(IRF-DMRG) method is developed for higher integer spin chain models 
which are rotational invariant.  The expressions of the IRF weights 
associated with the nearest-neighbor spin-$S$ interaction ${\bf S}_{i} 
\cdot {\bf S}_{i+1}$ are explicitly derived.  Using the IRF-DMRG with 
these IRF weights, the Haldane gaps $\Delta$ and the ground state 
energy densities $e_{0}$ for both $S=1$ and $S=2$ isotropic 
antiferromagnetic Heisenberg quantum spin chains are calculated by 
keeping up to only $m=90$ states.
\end{abstract}
 ]
 
\section{Introduction}
Density~matrix~renormalization~group~(DMRG) method has been a powerful 
numerical tool since White's pioneering works \cite{White}.  DMRG is a 
real space numerical renormalization method by using the reduced 
density matrix of a large system to approximate the ground (or 
excited) state(s) of the system with great accuracy.  There are
many DMRG applications for (quasi)one-dimensional (1D) systems,
e.g., quantum systems, 
statistical systems, polymers, etc.(for a review see 
Ref. \cite{Dresden Workshop}, and references therein).

DMRG is also a variational method and closely related to matrix 
product (MP) method \cite{MP,MP ladders,Equivalence VMP & 
DMRG,overview QG CFT}.  In MP method the ground state of a system, which is 
assumed to be expressed as a matrix product, can be obtained by 
minimizing the associated ground state energy density with respect to 
variational parameters.  On the other hand DMRG method may find
 the ground state of a large system, which consists of Wilsonian blocks,
 $B^L \bullet \bullet B^R$, by keeping the most probable $m$ states for each 
block $B^{L/R}$.  The most $m$ probable states are selected by 
using the $m$-largest eigenvalues of the reduced density matrix 
$\rho^{L/R}$ for each block.  For an overview and some connections to 
related fields including MP method, Ref.  \cite{overview QG CFT} 
is recommended.

When a system of interest has a symmetry, the associated eigenvalues of 
the density matrix $\rho$ are degenerated and one should keep the 
all states that corresponding to the same eigenvalues.  For example, if 
we consider a rotational invariant model, e.g., 1D isotropic Heisenberg 
spin models, with the standard vertex-DMRG, we may use third 
components $s_{z}$ of spin as basis, then the 
eigenvalues of the density matrix are degenerated due to the 
rotational symmetry.  One thus needs to keep many states to improve 
numerical accuracy with $s_{z}$ basis in order to get a real physics 
at thermodynamic limit.  There have been large-scale DMRG calculations 
with many states kept, for example, up to $m = 300$ \cite{S=2 Haldane 
gap}, $m = 400$ \cite{S=2 Haldane gap revisited}, $m=1700$ \cite{S=2 
Haldane gap Qin}, to estimate the $S=2$ Haldane gap.  However the more 
$m$ states are kept, the more computational cost is necessary since the 
dimension of the Hilbert space for the superblock $B^{L} \bullet 
\bullet B^{R}$ is proportional to $m^{2}$.

Sierra and Nishino\cite{IRF-DMRG} had applied the DMRG method to 
interaction round a face (IRF) models and developed the IRF-DMRG 
method. The advantage of using IRF-DMRG method in a system which
 has rotational symmetry is
 that the dimensions of the associated Hilbert 
spaces are much smaller than that in standard vertex-DMRG, since the 
degeneracy due to the symmetry has been eliminated.  They had 
demonstrated the power of the IRF-DMRG by calculating the ground state 
energies of the solid on solid (SOS) model, which is equivalent to 
spin-$1/2$ Heisenberg chain, and that of the restricted SOS (RSOS) 
model, which is equivalent to the quantum group invariant {\it XXZ} chain.  
They also had suggested a promising potential of the IRF-DMRG when it 
applies to higher integer quantum spin chains.  Such a work has not yet been 
done as far as I know.

In this work, the IRF-DMRG method is reviewed and further developed for the 1D 
integer spin antiferromagnetic Heisenberg (AFH) quantum spin chains.  
In IRF formulation the dynamics of a model can be described by a local 
plaquette operator $X_{i}$, which operates to the $i$th site of an 
IRF state as a ``diagonal to diagonal transfer matrix'' and their 
matrix elements are called {\it the IRF weights} as will be explained 
in the next section.  We hence need the explicit expressions of the 
IRF weights for the higher integer quantum spin chain models in order to work 
with the IRF-DMRG.

The rest of the paper is organized as follows.  In the next section 
the IRF-DMRG method is reviewed.  After the explanation of IRF 
formulation, the infinite system algorithm of the IRF-DMRG is discussed.  How 
to target the excited states of AFH quantum spin chains are explained 
and the superblock configuration suited for targeting the excited 
state of the AFH spin chain is proposed.  In Sec. III, the power of 
the IRF-DMRG is demonstrated by applying it to both $S=1$ and $S=2$ 
AFH quantum spin chains.  The Haldane gaps are calculated by keeping 
moderate number of $m$ states.  Finally conclusions are stated.  With 
the help of the Wigner Ekart's theorem summarized in Appendix A, I've 
derived the expression of the IRF weights for a nearest neighbor 
spin-$S$ interaction ${\bf S}_{i} \cdot {\bf S}_{i+1}$ in Appendix B. 
These expressions enable us to work with the IRF-DMRG.
\begin{figure}
   \begin{center}
        \epsfig{file=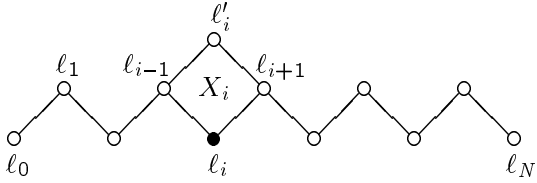,width=2.8in}\\
   \end{center}
   \caption{Diagrammatic representation of the operation of a plaquette 
   operator $X_{i}$ on a IRF state $|\ell_{0}, \cdots, \ell_{i}, \cdots, 
   \ell_{N} \rangle$, which is drawn as a zigzag chain of the lattice sites 
   (open circles). The closed circle stands for the summation over 
   the lattice variable $\ell_{i}$.  }
\label{plaquette diagram}
\end{figure}

\section{Review of IRF-DMRG}
The IRF-DMRG method is briefly reviewed here according to the Sierra 
and Nishino's paper \cite{IRF-DMRG}.  It is instructive to begin with 
a brief review of graph IRF model in somewhat general context.  The IRF 
representation of a rotational invariant quantum spin chain is then 
explained with the help of the associated spin graphs.  In fact 1D 
quantum spin chain models can be formulated as a special case of graph IRF 
models.  The algorithm of the IRF-DMRG is also reviewed.  A method how 
to target the excited states of AFH quantum spin chains and the appropriate 
superblock configurations are explained.

\subsection{Graph IRF model}
In interaction round a face or face models for short \cite{Flocke97,Baxter82}, 
a state is represented with the lattice variables $\ell$ assigned to 
the lattice sites;
\begin{equation}
  |{\bf \ell} \rangle = |\ell_{0}, \ell_{1},\ldots,\ell_{N} \rangle,
  \label{IRF-reps}
\end{equation}
\noindent
while the associated interaction is defined on a face, or plaquette 
of the lattice sites;
\begin{eqnarray}
  X_{i} | \dots, \ell_{i-1}, \ell'_{i}, \ell_{i+1}, \dots \rangle
  \hspace{1in} \nonumber \\
  = \sum_{\ell_{i}} R \left(
    \begin{array}{ccc}
      & \ell'_{i} & \\
     \ell_{i-1} & & \ell_{i+1} \\
      & \ell_{i} & \\
    \end{array}
  \right ) | \dots, \ell_{i-1}, \ell_{i}, \ell_{i+1}, \dots \rangle,
        \label{eq: plaquette}
\end{eqnarray}
\noindent where $R$ are so called {\it IRF weights}, which play an 
important role such as Boltzmann weights in usual statistical models.  The 
dynamics of the model is described by a plaquette operator $X_{i}$, 
which consists of the associated IRF weights.  The plaquette operator 
$X_{i}$ can be viewed as a local ``diagonal transfer matrix'' as shown 
in Fig. \ref{plaquette diagram} in contrast with a usual ``row to row 
transfer matrix.''

IRF model is characterized with a selection rule which determines 
whether the adjacent lattice sites of a lattice site are admissible or 
not.  The selection rule can be described by its associated incident 
matrix $\Lambda_{\ell,\ell'}$, whose elements  are all either 
$0$ or $1$.  If $\Lambda_{\ell,\ell'}=0$, then the lattice variables $\ell$ 
and $\ell'$ cannot be assigned to adjacent lattice sites, i.e., 
not admissible to each other.  The $\Lambda$ characterizes the set of all 
possible configurations which contribute to the Hamiltonian or 
partition function of the IRF model. A configuration not derived from 
$\Lambda$ necessarily has zero IRF weight.  It may be more convenient 
to use a graph or diagram associated with the selection rule, or 
$\Lambda$, of an IRF model and this is called a {\it graph} IRF (or 
graph face) model.  Each vertex of such a graph represents a lattice 
site $\ell_{i}$, which may take values from $1$ to $n_{i}$, and they 
are connected with line if they are admissible.
\begin{figure}
   \begin{center}
        \epsfig{file=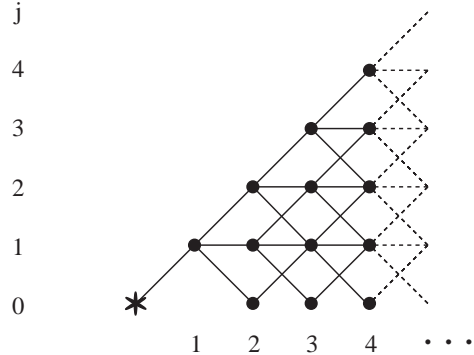,width=2.8in}\\
   \end{center}
   \caption{Spin diagram for $S=1$ chain.  Each vertex represents 
        an IRF lattice site.  The height of each site represents the 
        total spin angular momentum $j_{i}$ summing from the most left 
        vertex $\ast$ (vacuum state) until the $i$th spin.  
        Admissible vertices are connected with lines.}
\label{spin diagram}
\end{figure}
For a rotational invariant spin chain, the corresponding selection 
rule is nothing but the {\it addition rule} of spin angular momenta, 
each spin state is hence classified with the total spin angular 
momentum $j$.  The states of the chain with $N$ spins will be given by 
the set $|j_{N},M\rangle$ ($j^{\rm min} \le j_{N} \le j^{\rm max}$) of states 
with total spin $j_{N}$ and the corresponding third component $M$.  
One of the graphical representations associated with the addition rule 
of spin is the so called {\it spin diagram}.  For example, the spin 
diagram for $S=1$ spin chain is shown in Fig.  \ref{spin diagram}.  The 
corresponding lattice variables $\ell_{i}$ of the spin diagram are the 
magnitudes $j_{i}$ of spin angular momenta until the $i$th spin.  The 
height of each vertex represents the sum $j_i$ of the spin angular 
momenta starting from the most left spin until the $i$th spin.  The 
$i$th and ($i+1$)-th vertices are connected if they are admissible, 
i.e., if they are satisfied with the addition rule; $|j_i - S| \le 
j_{i+1} \le j_i + S$, where $S=1$ for spin-$1$ chain.  Note that $j_i$ 
takes values from $j_{i}^{\rm min}$ to $j_{i}^{\rm max}$, i.e., the range of 
$j_{i}$ is site dependent.  An IRF state of the $N$ spin chain is 
represented by a path in the spin diagram $|\ast, j_{1}, j_{2}, \cdots 
, j_{i}, \cdots , j_{N} \rangle$ and the most left vertex $\ast$ is 
corresponding to a vacuum state $|0 \rangle$.  If we know the all IRF 
weights associated with a spin chain model, we can work in the IRF 
formulation.  It is hence important to derive the explicit expressions 
for the IRF weights of the spin chain model to apply the IRF-DMRG. In 
Appendix B, I have derived the expression of the IRF weights for the 
nearest neighbor spin-$S$ interaction and summarized the diagrammatic 
representations and the corresponding IRF weights as a function of spin $j$
for the $S=1$ isotropic AFH case in Fig.  \ref{spin-1 R}.
\subsection{Algorithm}
The DMRG is an iterative algorithm to approximate a target 
state (the ground or excited state) of a large system which consists 
of the two blocks and lattice site(s).  In each iteration each block 
is extended by adding one lattice site but keeping only the most $m$ probable 
states, which are selected with the density matrix constructed from 
the target state. The superblock\cite{block} is formed by the left 
block $B^{L}$, the middle block which consists of a lattice point 
$\bullet$ or of two points $\bullet \bullet$, and the right block 
$B^{R}$, which is not necessarily the reflection of $B^{L}$.
\begin{figure}[t]
        \begin{center}
          \epsfig{file=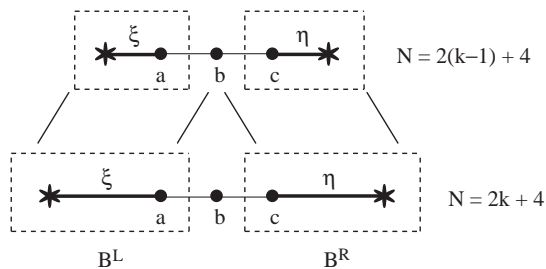,width=2.8in}\\
        \end{center}
        \caption{Schematic diagram of the infinite IRF-DMRG algorithm.  
        Each block (dotted line box) is extended by adding single 
        middle lattice point $\bullet$ in each DMRG iteration. 
        Thick lines represent renormalized block states.}
        \label{extension}
\end{figure}

Let me first review the infinite system algorithm \cite{IRF-DMRG} of 
the IRF-DMRG.  A Hilbert space of the superblock 
$B^{L} \bullet B^{R}$ can be written as
\begin{equation}
  {\mathcal H}^{B^{L} \bullet B^{R}} = \{ | \xi_{a} \otimes b 
  \otimes \eta_{c} \rangle \quad | \quad \Lambda_{a,b} = 
  \Lambda_{b,c} = 1 \},
  \label{H-space}
\end{equation}
\noindent i.e., a space spanned by a left block state $| \xi_{a} 
\rangle$, a middle lattice point state $| b \rangle$, and a right 
block state $| \eta_{c} \rangle$ (see Fig.  \ref{extension}), where 
the three lattice points (a, b, c) must be connected in a associated 
spin diagram. These lattice points take values from $j_{i}^{\rm min}$ to 
$j_{i}^{\rm max}$, where $i=a,b,c$.
\begin{figure}
        \begin{center}
          \epsfig{file=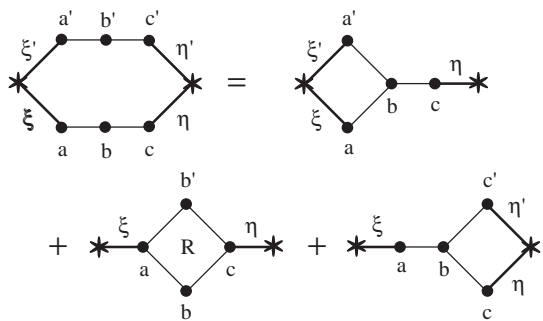,width=2.8in}\\
        \end{center}
        \caption{Diagrammatic representation for the construction of the super 
        block Hamiltonian $H^{B_{L} \bullet B_{R}}$; each diagram is 
        corresponding to each term in Eq. (\ref{H_SB}).}
        \label{3 block H}
\end{figure}
The superblock Hamiltonian $H^{B^{L} \bullet B^{R}}$ can be 
constructed from the two block Hamiltonians $H^{B^{L}}$, $H^{B^{R}}$ and
the IRF weights $R$ using the following relation,
\begin{eqnarray}
  H_{\xi_{a}, b, \eta_{c}}^{\xi'_{a}, b', \eta'_{c}}
  \left(
    \begin{array}{lllll}
      & a' & b' & c' &\\
      \ast & & & & \ast \\
      & a & b & c &
    \end{array}
  \right) = H_{\xi_a}^{\xi'_a} \left(
        \begin{array}{lll}
          & a' &\\
          \ast & & b \\
          & a &
        \end{array}
        \right)
        \delta_b^{b'} \Lambda_{b, c} 
        \delta_{\eta_c}^{\eta'_c} \nonumber \\
    + \delta_{\xi_a}^{\xi'_a} R \left(
      \begin{array}{lll}
        & b' &\\
        a & & c \\
        & b &
      \end{array}
        \right)
        \delta_{\eta_c}^{\eta'_c}
        + \delta_{\xi_a}^{\xi'_a} 
        \delta_{b}^{b'} \Lambda_{a, b} H_{\eta_c}^{\eta'_c} 
        \left(
        \begin{array}{lll}
          & c' &\\
          b & & \ast \\
          & c &
        \end{array}
        \right).
        \label{H_SB}
\end{eqnarray}
\noindent
Figure \ref{3 block H} shows the diagrammatic representation of the 
above equation.

We then may find the ground (or excited) state of 
the superblock by using a Lanczos or similar method,
\begin{equation}
  | \Psi_{G} \rangle = \sum_{\xi_{a}, b, \eta_{c}} 
  \psi^{b}_{\xi_{a}, \eta_{c}} | \xi_{a} \otimes b \otimes \eta_{c} \rangle.
  \label{Psi_G}
\end{equation}
\noindent
The left density matrix $\rho^{B^{L} \bullet}$ is readily
constructed as
\begin{equation}
        \rho^{\xi'_{a}}_{\xi_{a}} \left(
        \begin{array}{lll}
                & a' &\\
            \ast & & b \\
            & a &\\
        \end{array}
        \right) = \sum_{\eta_{c}} 
        \psi^{b}_{\xi'_{a}, \eta_{c}} \psi^{b}_{\xi_{a}, \eta_{c}}.
        \label{rho_L}
\end{equation}
\noindent
The right density matrix $\rho^{B^{R} \bullet}$ is constructed in the 
similar way if necessary.

The next step is to diagonalize $\rho^{B^{L} \bullet}$ to obtain the 
eigenvalues $w^{b}$ and the associated eigenvectors $|u^{b}_{\xi_{a}} 
\rangle$.  Keeping the $m$ states associated with $m$-largest $w^{b}$ 
($m = \sum_{b} m_{b}$) in order to form the projection operator 
$T^{\dag} T$.
\begin{equation}
        T = \sum_{b} T^{b}, \hspace{1cm} T^{b} = \sum_{\xi_{a}} 
        |u^{b}_{\xi_{a}} \rangle \langle u^{b}_{\xi_{a}} |
        \label{eq:T}
\end{equation}
\noindent
The operator $T$ truncates the Hilbert space ${\mathcal 
H}^{B^{L} \bullet}$ into ${\mathcal H}^{B'^{L}}$, where $B'^{L}$ 
represents a block with one more lattice site than the block $B^{L}$.  
A new left block Hamiltonian is then formed by using the projection 
operator as $H^{B'^{L}} = T (H^{B^{L} \bullet}) T^{\dag}$.  In each 
IRF-DMRG iteration the system is extended by adding single middle site 
$\bullet$ to both blocks as shown in Fig.  \ref{extension}.

The infinite system algorithm for the IRF-DMRG with the superblock $B^{L} 
\bullet B^{R}$ is summarized as follows:
\def\labelenumi{\theenumi)}
\def\theenumi{\roman{enumi}}
\begin{enumerate}
\item find the ground state of the superblock Hamiltonian, Eq.  
  (\ref{H_SB});
\item form the left (right) reduced density matrix 
  $\rho^{B^{L} \bullet}$ ($\rho^{B^{R} \bullet}$), Eq.  
  (\ref{rho_L});
\item diagonalize $\rho^{B^{L} \bullet}$ 
  ($\rho^{B^{R} \bullet}$) to obtain the eigenvalues $w^{b}$ and the 
  associated eigenvectors $|u^{b}_{\xi_{a}} \rangle$ ( $|u^{b}_{\eta_{c}} 
  \rangle$);
\item keep the 
  $m$ states $|u^{b}_{\xi_{a}} \rangle$ corresponding to the $m$ 
  largest eigenvectors $w^{b}$ to form the projection operator $T$, 
  Eq. (\ref{eq:T});
\item renormalize the block operators by using $T$: \\
  $H^{B'^{L}} = T (H^{B^{L} \bullet}) T^{\dag}$, etc.;
\item extend each block by adding one site $\bullet$;
\end{enumerate}
\noindent
repeat the processes i)-vi) for new blocks.
\begin{figure}
        \begin{center}
                \epsfig{file=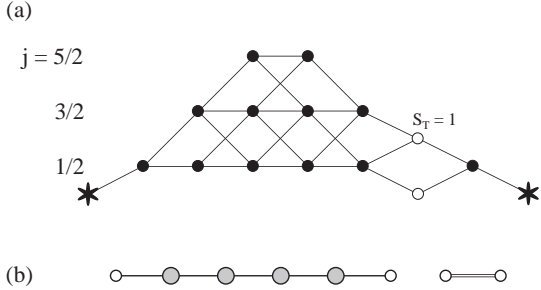,width=2.8in}\\
        \end{center}
        \caption{(a) Spin diagram of the superblock for targeting the 
        excited state or the ground state of the $S=1$ open spin chain 
        ended with $S=1/2$ spins.  (b) The corresponding spin chain of 
        $N=4$ spin-$1$ spins (large hatched circles) ended with 
        spin-$1/2$ spins (small open circles).  For targeting the excited 
        states, which has total spin $S^{T}=1$, ferromagnetic coupled two 
        spin-$1/2$ spins are attached.}
        \label{spin-1 diagram}
\end{figure}
\begin{figure}
        \begin{center}
          \epsfig{file=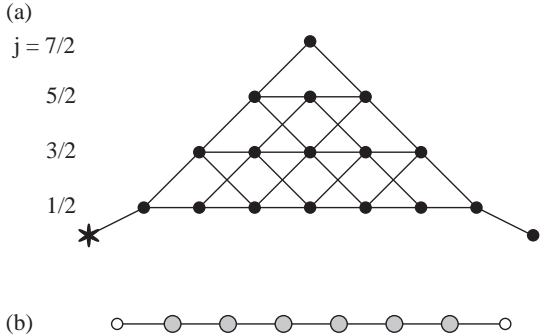,width=2.8in}\\
        \end{center}
        \caption{(a) Spin diagram of $S=1$ chain ended with $S=1/2$ spins.  
        AKLT state is corresponding to the horizontal line at $j=1/2$.  
        (b) The corresponding spin chain of $N=6$ spin-$1$ spins (large 
        hatched circles) ended with spin-$1/2$ spins (small open circles).}
        \label{spin-1 GS diagram}
\end{figure}
\begin{figure}
        \begin{center}
          \epsfig{file=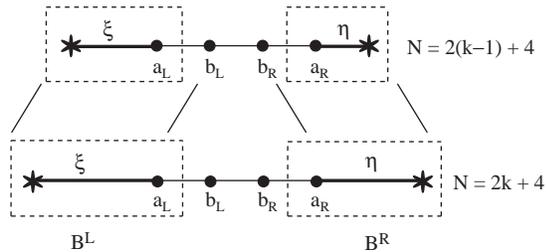,width=2.8in}\\
        \end{center}
        \caption{Schematic diagram of the proposed infinite IRF-DMRG algorithm.
        The left (right) block is extended by adding single middle left 
        (right) point $\bullet$ in each 
        IRF-DMRG iteration}
        \label{BssB extension}
\end{figure}
Next let me explain how to target the excited state of the AFH spin chain 
before discussing the superblock configuration suited for the excited state.
Figure  \ref{spin-1 diagram} shows how to target the 
excited state of the AFH spin-$1$ spin chain ended with spin-$1/2$ 
spins.  In order to fix the total spin $S_{T}$ of the chain in 
$S_{T}=1$ sector, two spin-$1/2$ spins that ferromagnetically coupled with 
the coupling constant $J_{F}<0$ are attached as shown in Fig.  
\ref{spin-1 diagram} (b).  Note that the attached two spins, which 
energetically favor a triplet state, are not coupled with the chains.  
Since the whole system (the chain and the two spins) is set to a 
singlet state ($0$ sector), the triplet state of the two spins enforce 
$S_{T}=1$ on the chain \cite{targetting excited state}.

Now let us consider the superblock configuration that suited for targeting 
the excited state.  In the infinite system algorithm of the IRF-DMRG with the 
superblock configuration of $B^{L} \bullet B^{R}$, the middle point is 
single {\it bare site} and plays an important role to successively 
improve the renormalized block Hamiltonians.  In each IRF-DMRG 
iteration both blocks are renormalized with the middle site as shown in 
Fig.  \ref{extension}.  In Ref.  \cite{IRF-DMRG} the superblock 
configuration of $B^{L} \bullet B^{R}$ was used and it is suited for 
targeting the ground states of AFH spin chains, because 
the highest $j$ site always lies in the middle of the superblock as 
shown in Fig.  \ref{spin-1 GS diagram}.  Since the ground state of an 
AFH quantum spin chain with even $N$ spins lies in the total spin $0$ 
sector, $j_{N}=0$ and obviously the vacuum state $\ast$ has zero spin, 
i.e., $j_{0}=0$.  Furthermore the system is reflection symmetric with 
respect to the middle point in the spin diagram.  Thus the middle 
point always has the highest $j$ in the IRF-DMRG formulation when 
targeting the ground state of the AFH spin chain.  In other word the 
highest $j$ of the left block, which is constructed from the left 
vacuum by adding spins, coincides with that of the right block, which 
is constructed from the right vacuum, at the middle point in the spin 
graph.  
\begin{figure}
        \begin{center}
                \epsfig{file=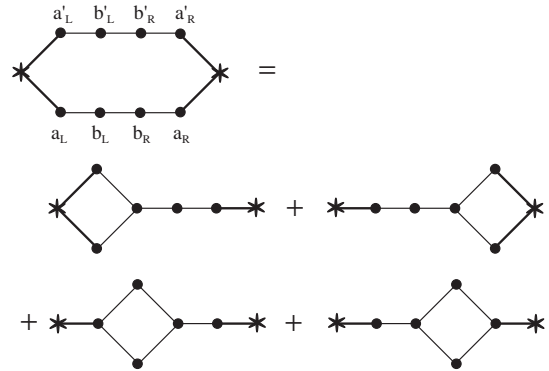,width=2.8in}\\
        \end{center}
        \caption{Diagrammatic representation for the construction of the super 
        block Hamiltonian $H^{B_{L}\bullet \bullet B_{R}}$.}
        \label{4 block H}
\end{figure}
However the situation is somewhat different for targeting the 
excited state which lies in $S_{T}=1$ sector.  From Fig.  \ref{spin-1 
diagram} (a), we noticed that no single middle point at which the 
highest $j$ of the left block coincides with that of the right one and 
that there is no reflection symmetry.  Hence it is better to use the 
superblock configuration of $B^{L} \bullet \bullet B^{R}$, at least 
when we use the infinite system method, and 
renormalize $B^{L}$ ($B^{R}$) with the middle left (right) site 
individually as shown in Fig.  \ref{BssB extension}.  The associated 
superblock Hamiltonian can be 
constructed in the way diagrammatically shown in Fig.  \ref{4 block 
H}.

\section{Application to $S=1$ and $S=2$ antiferromagnetic Heisenberg 
quantum spin chains}
The supremity of the IRF-DMRG method is demonstrated by applying 
to both $S=1$ and $S=2$ quantum spin chains.  Haldane's conjecture 
\cite{Haldane} that the physics of isotropic antiferromagnetic quantum 
spin chains depend substantially on whether the spin is integer or 
half-integer, has been motivating many physicists to study quantum 
spin chains.  $S=1$ AFH chain have been widely studied with various 
methods \cite{S=1 chains}.  It is a well known fact that the ground 
state of an open $S=1$ quantum 
spin chain has an effective $S=1/2$ spin at each end.  
White \cite{S=1 Heisenberg} had obtained the ground state energy per 
site of $e_{0} \cong -1.401484038971(4)$ with $m = 180$ states kept 
and the Haldane gap of $\Delta_{1} \cong 0.41050(2)$ with $m = 160$ 
states kept by using the standard $s_{z}$-base DMRG.
\begin{figure}
        \begin{center}
                \epsfig{file=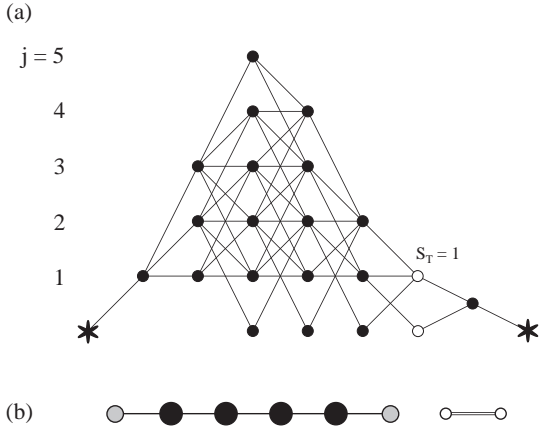,width=2.8in}\\
        \end{center}
    \caption{(a) Spin diagram of open $S=2$ spin chain ended with $S=1$ spins.
    The total $S^T=1$ state is targeted to find the excited state. 
    (b) The corresponding spin chain of $N=4$ spin-$2$ spins 
    (large closed circles) ended with spin-$1$ spins (hatched circles). 
    The attached two spin-$1/2$ spins (small open circles) are coupled 
    ferromagnetic or antiferromagnetic depending on the target state.}
    \label{spin-2 diagram}
\end{figure}
$S=2$ AFH quantum spin chains have also been studied \cite{S=2 Haldane 
gap,S=2 Haldane gap revisited}, although the numerical calculations 
are much elaborative than that of $S=1$ case due to the longer 
correlation length $\xi$ and due to the larger number of degrees of 
freedom per spin because the degeneracy due to the spin symmetry.  The 
longer correlation length means that much longer chains (more than one 
thousand) are required to reach the convergence regime to exclude the 
finite size effects.  Schollw\"{o}ck {\it et al}.\cite{S=2 Haldane 
gap} estimated that the gap of $\Delta_{2} \cong 0.085(5)$ and the 
ground state energy density to be $e_{0} \cong -4.761248(1)$ by using 
the standard DMRG with up to $m=210$ states kept.
Wang {\it et al}.  \cite{S=2 Haldane gap revisited} systematically 
analyzed and estimated that the gap $\Delta_{2} \cong 0.0876 \pm 0.0013$ 
with up to $m=400$ states kept during DMRG calculations.

Following Schollw\"{o}ck {\it et al}., 
I also consider the open spin-$S$ AFH quantum spin chain 
terminated with spin-$S/2$ spins to cancel out the effective spin-$S/2$ 
spins:
\begin{equation}
    H = J_{\rm end} {\bf S}_{1} \cdot {\bf S}_{2} + J \sum_{i=2}^{N-2}
    {\bf S}_{i} \cdot {\bf S}_{i+1} + J_{\rm end} {\bf S}_{N-1} \cdot {\bf S}_{N},
\end{equation}
\noindent 
where the both coupling constants $J$ and $J_{\rm end}$ are set 
to unity for simplicity.  The spin diagram for the $S=1$ ground state 
is shown in Fig.  \ref{spin-1 GS diagram}.  I have obtained the 
comparable result of $e_{0} \cong -1.40148403897$ with the IRF-DMRG by 
keeping only $m=80$ states, which consist of $25 (j=1/2)$, $31 
(j=3/2)$, $19 (j=5/2)$, and $5 (j=7/2)$ $m_{j}$ states, without resorting 
to a scaling technique.  Note that since $m_{j}$ states in $j$ base 
correspond to $(2j+1) m_{j}$ in $s_{z}$ base, the above $m=80$ 
correspond to $m=328$ in $s_{z}$ base.  For the $S=2$ chain with $N 
\approx 200$, I have found that the ground state energy density is 
$e_{0} \cong -4.7612481(6)$ by keeping $m=90$ states, which consist of 
$9 (j=0)$, $24 (j=1)$, $27 (j=2)$, $19 (j=3)$, $9 (j=4)$, and $2 
(j=5)$ $m_{j}$ states, hence corresponding to $m=452$ states in 
$s_{z}$ base.
 
\begin{figure}
        \begin{center}
          \epsfig{file=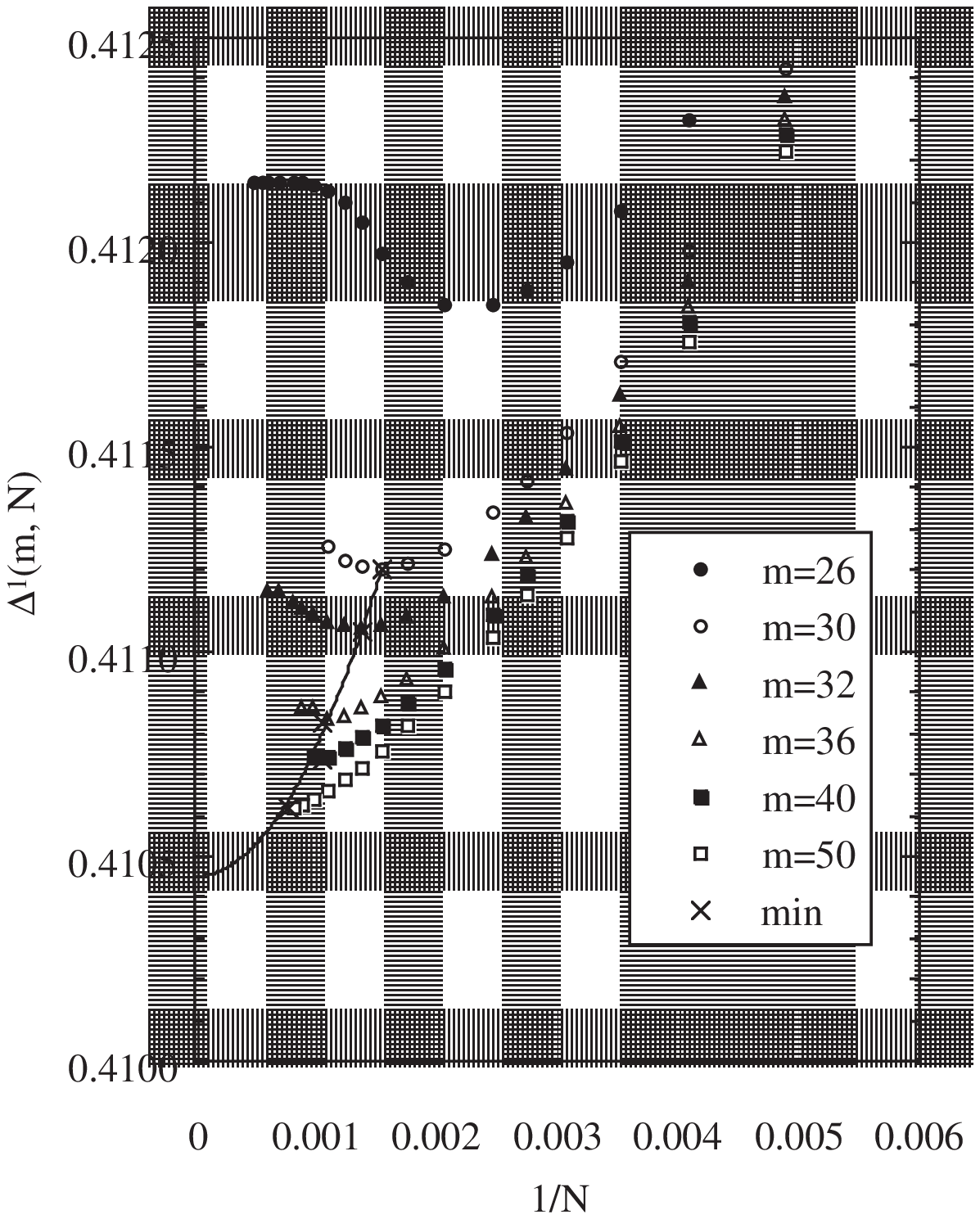,width=2.8in}\\
        \end{center}
    \caption{The gap $\Delta^{1}(m, N)$ in the unit of 
    the coupling constant $J$ as a function of the number of 
    state $m = \sum_{j} m_{j}$ kept in the IRF-DMRG calculations and 
    the spin chain length $N$.
    Each cross denotes the position of minimum for a given $m$ curve.}
        \label{Egap spin-1}
\end{figure}

The finite size correction to the Haldane gap $\Delta_{S}$ with open 
boundary conditions is proportional to the inverse of the square chain 
length according to the 1D field theory \cite{AKLT,finite size correction}:
\begin{equation}
        \Delta(m = \infty, N) = \Delta_{S} + \frac{v^{2} \pi^{2}}{2 
        \Delta_{S} N^{2}} + {\mathcal O}(\frac{1}{N^{3}}),
        \label{scaling}
\end{equation}
\noindent where $v$ is the spin wave velocity and $\Delta_{S}$ the 
Haldane gap of the spin-$S$ AFH spin chain at the thermodynamic limit.  
Figure  \ref {Egap spin-1} indicates the gap $ \Delta_{1}(m, N)$ as 
measured by the difference between the lowest energy of $S^{T}=1$ 
states and that of $S^{T}=0$ states, as a function of the spin chain 
length $N$ and the number of states $m$ kept in the IRF-DMRG 
iterations.  The excited states were obtained by using the method 
explained in the previous section as shown in Figs. \ref{spin-1 diagram}
 and \ref{BssB extension}.  With the simultaneous 
extrapolation for $m$ and $1/N$, the Haldane gap for the $S=1$ AFH 
chain is estimated to be $\Delta_{1} \cong 0.4104(5)$.

\begin{figure}
        \begin{center}
          \epsfig{file=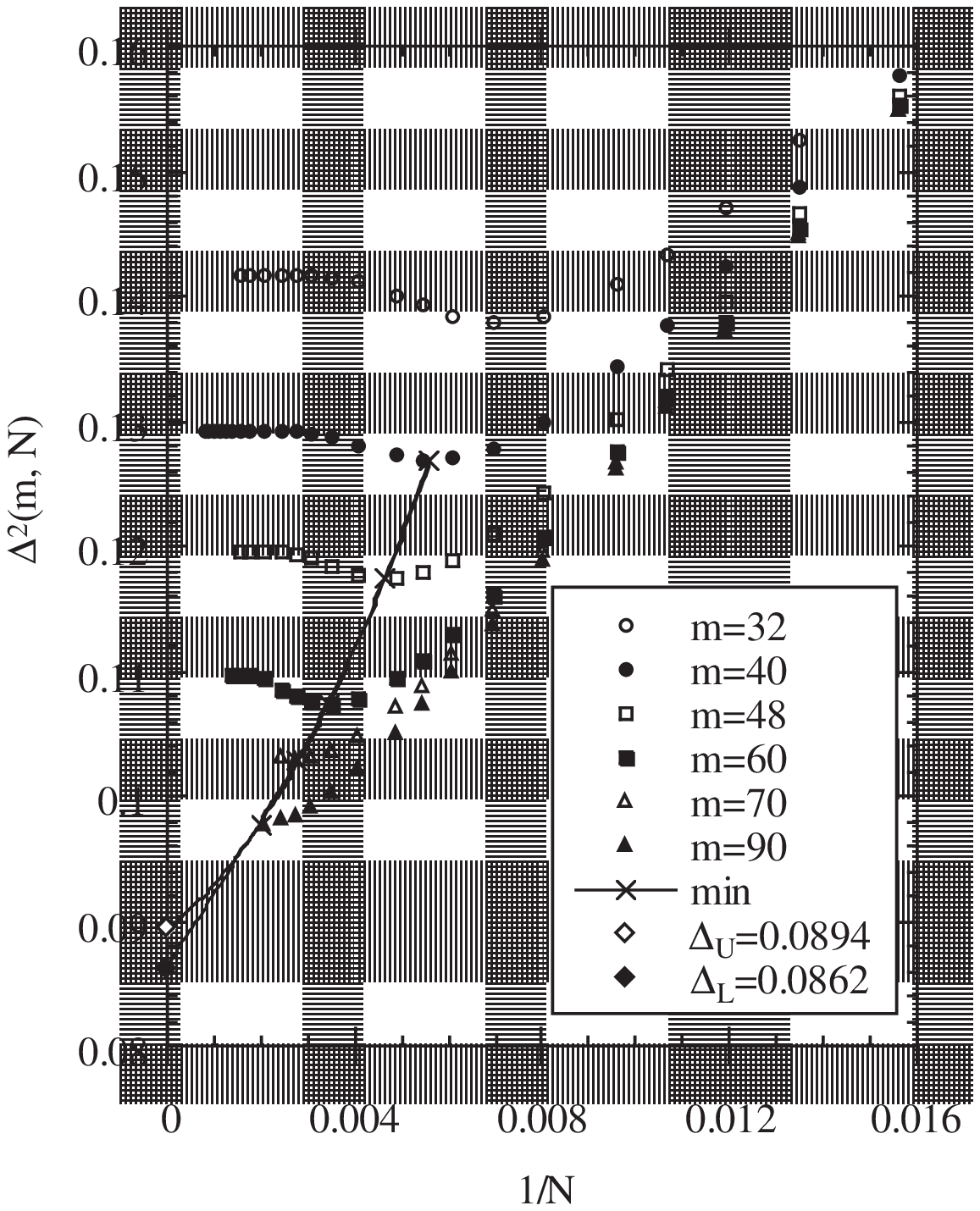,width=2.8in}\\
        \end{center}
        \caption{The gap $\Delta^{2}(m, N)$ in the unit of the coupling
        constant $J$ as a function 
        of the number of 
        state $m = \sum_{j} m_{j}$ kept in the IRF-DMRG calculations and 
        the spin chain length $N$.  Each cross denotes the position of 
        minimum for a given $m$ curve.  The open and solid diamonds denote 
        upper estimated value $\Delta_{U}$ and lower one $\Delta_{L}$, 
        respectively.}
        \label{Egap spin-2}
\end{figure}

Wang {\it et al}.  \cite{S=2 Haldane gap revisited} pointed out that one 
cannot use the extrapolation with respect to $1/N$ to obtain the gap 
value at the thermodynamic limit when $m$ is not sufficiently large: 
the minimum of each curve for $\Delta_{S}(m, N)$ deviates from the 
vertical axis as $m$ decreases; while Eq.  (\ref{scaling}) tells us the 
minimum should be located just on the vertical axis in the limit $m 
\to \infty$.  DMRG involves a systematic error associated with the 
truncation or keeping a finite $m$ states of the renormalized Hilbert 
space.  They thus remarked that these errors are not so serious for 
$S=1/2$ or $S=1$ AFH chains but the errors become crucial for higher 
spin chains and scaling for $m$ should be carefully carried out.  In 
standard DMRG calculations the multiplicity $2S+1$ of spin-$S$ is not 
eliminated.  Then in order to treat the higher spin chain, 
the larger number $m$ of 
states should be kept.  The IRF-DMRG however has a great merit to 
eliminate the degeneracy due to the spin symmetry.  We can hence keep 
effectively larger $m$ states!
Figure  \ref{Egap spin-2} shows the gap $\Delta_{2}(m, N)$, as measured 
by the difference between the lowest energy of $S^{T}=1$ states and 
that of $S^{T}=0$ states, as a function of the spin chain length $N$ 
and the number of states $m$ kept in the IRF-DMRG iterations.  The 
extrapolations are performed for both $m$ and $N$ simultaneously and 
the upper estimated value $\Delta_{U}$ and lower one $\Delta_{L}$ are 
obtained by using the polynomial fits with second order and third 
order, respectively.  The estimated Haldane gap for the $S=2$ AFH chain is 
$\Delta_{2} \cong 0.0878 \pm 0.0016$.

\section{Conclusions}
The IRF-DMRG is reviewed and developed for higher integer quantum spin 
chain models which has a rotational symmetry.  The explicit 
expressions of the IRF weights for the nearest neighbor spin-$S$ 
interaction have been derived as a function of total spin $j$.  Using 
these IRF weights the IRF-DMRG has been applied to both $S=1$ and 
$S=2$ isotropic AFH quantum spin chains.  With the moderate number (up 
to $m=90$) of states kept in the IRF-DMRG iterations, the Haldane gaps 
and ground state energy densities were readily calculated since the 
degeneracy due to the spin symmetry had been eliminated.
\section*{Acknowledgments}
I thank Nishino Tomotoshi for valuable and 
helpful discussions mainly with the DMRG mailing list \cite{DMRG ML}.
I also acknowledge the organizers and participants of the workshop 
``DMRG '98'' held at Max 
Planck Institute f\"{u}r Physik Komplexer System in Dresden for hospitality
 and valuable discussions.  Most of 
the computations were carried out with newmat09 \cite{newmat} matrix 
library and Standard Template Library (STL) in C++ (egcs-1.1.2 
\cite{egcs}) on both Linux-Alpha (Stataboware 2.01$\beta$ 
\cite{stataboware}) and NetBSD-Alpha \cite{NetBSD} machines.  I 
further thank Goto Kazushige for providing a stable and powerful 
Linux-Alpha OS, Stataboware.
\appendix
\section{Wigner-Eckart's theorem}
The Wigner-Eckart's theorem is briefly reviewed to derive the IRF 
weights associated with general spin-$S$ interaction in the form of 
$\bf{S} \cdot \bf{S}'$ in the next appendix. Most of the results 
presented here are already known and can be found in Ref. \cite{MP ladders}.

Let $T_M^{(k)}$ is an irreducible tensor operator with an angular 
momentum ${\bf k}$ and the third component $M( = -k, \cdots, k)$.
$T_M^{(k)}$ commutes with the total 
angular momentum operator ${\bf J}$ of a system considered as
\begin{eqnarray}
        \lbrack J_z, T_M^{(k)} \rbrack & = & M T_M^{(k)} \\
        \lbrack J_x \pm i J_y, T_M^{(k)} \rbrack & = & 
\sqrt{k(k+1)-M(M\pm1)} 
        T_{M\pm1}^{(k)}.
\end{eqnarray}
\noindent i.e., $T_M^{(k)}$ is transformed as a tensor under a 
rotational operation for the system.  For example, each component of 
spin operator ${\bf S}$ is expressed as
\begin{eqnarray}
        T_0^{(1)} & = & S_z, \nonumber \\
        T_{\pm1}^{(1)} & = & \frac{\mp 1}{\sqrt{2}} S_{\pm}.
\end{eqnarray}

The inner product of a pair of irreducible tensors ${\bf T^{(k)}}$ 
and ${\bf U^{(k)}} $ is defined with
\begin{equation}
        {\bf T^{(k)}} \cdot {\bf U^{(k)}} \equiv \sum_{M=-k}^{k} 
        (-1)^{-M} T^{(k)}_{M} U^{(k)}_{-M}.
\end{equation}
\noindent For spin operator one readily checks the following relation:
\begin{equation}
        {\bf T^{(1)}} \cdot {\bf T^{(1)}} = {\bf S} \cdot {\bf S}.
\end{equation}

Wigner-Eckart's theorem shows the matrix elements for $T_M^{(k)}$ are 
factored as a product of a configuration dependent part, which 
expressed as Wigner's $3$-$j$ symbols or Clebsh-Gordan (CG) coefficients, 
and a configuration independent part:
\begin{eqnarray}
  \langle J M | T^{(k)}_{\mu} | J' M' \rangle \hspace{1.5in}
  \nonumber \\
  = (-1)^{J-M} \left[
    \begin{array}{ccc}
      J & k & J' \\
      -M & \mu & M'
    \end{array}
  \right] (J|| {\bf T}^{(k)} ||J'),
  \label{WE theorem}      
\end{eqnarray}
\noindent where $(J|| {\bf T}^{(k)} ||J')$ are called {\it reduced matrix 
elements}, which is independent of the configuration $M$ or $M'$.  
Wigner's $3$-$j$ symbol is related with CG coefficient as
\begin{equation}
  \left[
    \begin{array}{ccc}
      J & k & J' \\
      -M & \mu & M
    \end{array}
  \right] = \frac{(-1)^{J-k-M'}}{\sqrt{2J'+1}} \langle J -\!M k \mu 
  | J' -\!M' \rangle.
\end{equation}

Once the reduced matrix elements are known, one gets the all matrix 
elements with only calculations for CG coefficients.  For example one
easily finds the reduced matrix element for spin operator ${\bf S}$ by 
applying the theorem Eq. (\ref{WE theorem}) to $S^{z}$, 
which is corresponding to $k=1$ and $\mu=0$, as the following:
\begin{equation}
        \langle S M | T^{(1)}_{0} | S M' \rangle = (-1)^{S-M} \left[
    \begin{array}{ccc}
                S & 1 & S \\
       -M & 0 & M'
   \end{array}
   \right] (S|| {\bf S} ||S).
\end{equation}
\noindent Since the left hand side is $M \delta_{M, M'}$, one gets
\begin{equation}
   (S|| {\bf S} ||S) = \sqrt{S(S+1)(2S+1)}.
\end{equation}

The following relations \cite{MP ladders} are used to derive the 
expression for the IRF-weights of general spin-$S$ chain in the next 
appendix:
\begin{eqnarray}
        \langle j_1 j_2 J M | {\bf T}^{(k)}_{1} \cdot {\bf T}^{(k)}_{2} 
    | j'_1 j'_2 J' M' \rangle \qquad \qquad \nonumber \\
    = \delta_{J J'} \delta_{M M'} (-1)^{j_2+J+j'_1} \left\{
    \begin{array}{ccc}
                j_1 & j_2 & J \\
        j'_2 & j'_1 & k
    \end{array}
    \right\} \nonumber \\
    \qquad \times (j_1|| {\bf T}^{(k)}_1 ||j'_1) 
    (j_2|| {\bf T}^{(k)}_2 ||j'_2),
    \label{WE}
\end{eqnarray}
\begin{eqnarray}
        ( j_1 j_2 J || {\bf T}^{(k)}_{1} || j'_1 j'_2 J' ) \hspace{1 in}
    \qquad \nonumber \\
    = \delta_{j_2 j'_2} (-1)^{j_1+j_2+J'+k} \left\{
    \begin{array}{ccc}
                j_1 & J & j_2 \\
        J' & j'_1 & k
        \end{array}
        \right\} \nonumber \\
        \qquad \times \sqrt{(2J+1)(2J'+1)} (j_1|| {\bf T}^{(k)}_1 ||j'_1),
        \label{WE1}
\end{eqnarray}
\begin{eqnarray}
        ( j_1 j_2 J || {\bf T}^{(k)}_{2} || j'_1 j'_2 J' ) \hspace{1 in}
    \qquad\nonumber \\
    = \delta_{j_1 j'_1} (-1)^{j_1+j'_2+J+k} \left\{
    \begin{array}{ccc}
                j_2 & J & j_1 \\
                J' & j'_2 & k
        \end{array}
    \right\} \nonumber \\
    \qquad \times \sqrt{(2J+1)(2J'+1)} (j_2|| {\bf T}^{(k)}_2 ||j'_2).
    \label{WE2}
\end{eqnarray}
\section{IRF-weights}
I here derive the IRF weights for the following Hamiltonian $H$, 
which is invariant under rotations.
\begin{equation}
        H_{i, i+1} = {\bf S}_i \cdot {\bf S}_{i+1},
\end{equation}
\noindent
where ${\bf S}_i$ and ${\bf S}_{i+1}$ are not necessarily same.
The IRF weights for the $H$ are expressed as the matrix elements 
\begin{equation}
  \langle (J_i S_{i+1}), J_{i+1} M_{i+1} | {\bf S}_i \cdot {\bf S}_{i+1} | 
  (J'_i S_{i+1}), J'_{i+1} M'_{i+1} \rangle,
\end{equation}
\noindent
where $| (J_i S_{i+1}), J_{i+1} M_{i+1}\rangle$ is a state that the sum of 
spin angular momenta until the $(i+1)$-th spin is $J_{i+1}$ and the 
corresponding third component is $M_{i+1}$. The $J_{i+1}$ consists of 
$J_i$ and a spin-$S_{i+1}$ spin.

Using Eq. (\ref{WE}) one finds
\begin{eqnarray}
  \langle (J_i S_{i+1}), J_{i+1} M_{i+1} | {\bf S}_i \cdot {\bf S}_{i+1} | 
    (J'_i S_{i+1}), J_{i+1} M_{i+1} \rangle \nonumber \\
    = (-1)^{J_{i+1}+J_i+S_{i+1}}
    \left\{
      \begin{array}{ccc}
        J_i & S_{i+1} & J_{i+1} \\
        S_{i+1} & J'_i & 1 
      \end{array}
    \right\}  \nonumber \\
    \quad \times (S_{i+1} || {\bf S} || S_{i+1}) (J_i|| {\bf S} ||J'_i).
\end{eqnarray}
\noindent 
Since  $|J_i)$ can be expressed as a tensor product of $J_{i-1}$ and spin-$S$, the last reduced matrix element in the above equation may be written, with the help of Eq. (\ref{WE1}), as
\begin{eqnarray}
        (J_i|| {\bf S} ||J'_i) =
                ( J_{i-1} S_{i} J_i || {\bf S} || J'_{i-1} S_{i} J'_{i} )
                \hspace{0.8in} \nonumber \\
        = \delta_{J_{i-1} J'_{i-1}} (-1)^{J'_{i}+S_{i}+J_{i-1}+1}
                \left\{
                \begin{array}{ccc}
                        S_{i} & J_i & J_{i-1} \\
                        J'_{i} & S_{i} & 1
        \end{array}
        \right\} \nonumber \\
        \quad \times \sqrt{(2J_i+1)(2J'_i+1)} (S_{i} || {\bf S} || S_{i}).
\end{eqnarray}
\noindent
Hence the final expression is obtained as
\begin{eqnarray}
  R \left(
    \begin{array}{ccc}
      & J'_i &\\
      J_{i-1} & & J_{i+1} \\
      & J_i &
    \end{array}
  \right) \hspace{2 in} \nonumber \\
  \equiv \langle (J_i S_{i+1}), J_{i+1} | {\bf S}_i \cdot 
  {\bf S}_{i+1} | (J'_i S_{i+1}), J_{i+1} \rangle \qquad \nonumber \\
  = (-1)^{S_{i}+S_{i+1}+J_{i-1}+J_i+J'_i+J_{i+1}+1}
  \sqrt{(2J_i+1)(2J'_i+1)} \nonumber \\
  \times \sqrt{S_i(S_i+1)(2S_i+1)} 
  \sqrt{S_{i+1}(S_{i+1}+1)(2S_{i+1}+1)}\quad \nonumber \\
  \times \left\{
    \begin{array}{ccc}
      J_i &  S_{i+1}& J_{i+1} \\
      S_{i+1} & J'_i& 1
    \end{array}
  \right\} \left\{
    \begin{array}{ccc}
      S_{i} & J_i & J_{i-1} \\
      J'_{i} & S_{i} & 1
    \end{array}
  \right\}.
  \label{IRF weight}
\end{eqnarray}

\subsection{$S=1$ nearest-neighbor interaction}
\begin{figure}
 \begin{center}
       \epsfig{file=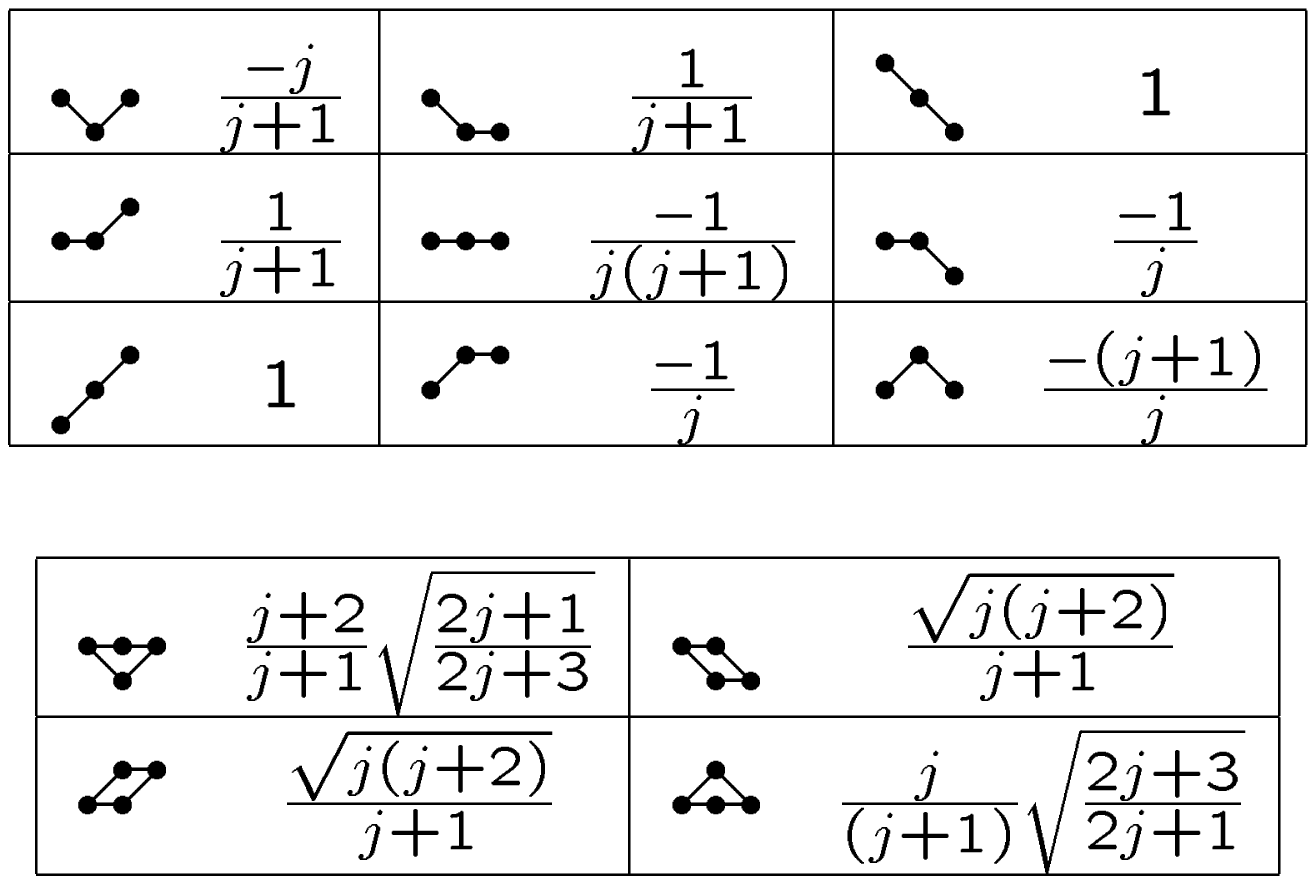,width=2.8in}\\
  \end{center}
         \caption{The diagrammatic representations and corresponding 
          expressions for the IRF-weights of 
          the $S=1$ nearest-neighbor interaction ${\bf S}_{i} \cdot 
         {\bf S}_{i+1}$.  In each diagram the hight of a vertex represents the 
         magnitude of the spin angular momentum assigned to the
         vertex.  The height of the (lower) middle vertex is assumed to be 
         $j$.}
 \label{spin-1 R}
 \end{figure}

There are thirteen nontrivial IRF weights R for the $S=1$ nearest 
neighbor spin interactions: ${\bf S}_{i} \cdot {\bf S}_{i+1}$.  They 
can be obtained by substituting $S_{i}=S_{i+1}=1$ in Eq.  (\ref{IRF 
weight}) and the results as a function of $j$ are summarized in Fig.  
\ref{spin-1 R} with diagrammatic representation.  In each diagram 
the magnitude of spin angular momentum for the middle vertex is 
assumed to be $j$, and 
the height of each vertex represents the magnitude of the spin angular
momentum that assigned to the vertex.  The first diagram, for example, 
stands for 
$ R \left(
        \begin{array}{ccc}
          & j  & \\
         j+1 & & j+1 \\
          & j & \\
        \end{array}
  \right ) = \frac{-j}{j+1}$.
\noindent
Similar results for $S=2$ case are readily obtained with the help of 
a symbolic manipulation language such as {\sc mathematica}.


\begin{references}
\bibitem{block}
In Ref. \cite{IRF-DMRG} the term ``super-string'' is used, but I use 
``superblock'' through out this paper.
\bibitem{stataboware}
A stable and robust Linux-Alpha package developed by Goto Kazushige 
for numerical computation.  It is still under development and 
unfortunately there is no detailed document but a brief README file in 
Japanese.
\end{references}
\end{document}